\documentclass[amsmath,amssymb,floatfix,twocolumn,superscriptaddress]{revtex4-2}

\usepackage{graphicx}
\usepackage{epsfig}
\usepackage{amsmath}
\usepackage{color}
\usepackage[colorlinks=true, letterpaper=true, pdfstartview=FitV, linkcolor=blue, citecolor=blue, urlcolor=blue]{hyperref}
\usepackage{ulem}

\setcitestyle{super,open={},close={}}
\renewcommand{\onlinecite}[1]{\nocite{#1}\citenum{#1}}
\makeatletter
\renewcommand\@biblabel[1]{#1.}
\makeatother

\begin{document}

\title{Correlated topological band structures of the kagome altermagnets Mn$_3X$ ($X=$ Sn, Ge, Ga)}

\author{Yingying Cao}
\affiliation{Beijing National Laboratory for Condensed Matter Physics and
Institute of Physics, Chinese Academy of Sciences, Beijing 100190, China}
\affiliation{School of Physical Sciences, University of Chinese Academy of Sciences, Beijing 100049, China}
\affiliation{Institute of Quantum Materials and Physics, Henan Academy of Sciences, Zhengzhou 450046, China}
\author{Yuanji Xu}
\affiliation{Institute for Applied Physics, University of Science and Technology Beijing, Beijing 100083, China}
\author{Yi-feng Yang}
\email[]{yifeng@iphy.ac.cn}
\affiliation{Beijing National Laboratory for Condensed Matter Physics and
Institute of Physics, Chinese Academy of Sciences, Beijing 100190, China}
\affiliation{School of Physical Sciences, University of Chinese Academy of Sciences, Beijing 100049, China}
\affiliation{Songshan Lake Materials Laboratory, Dongguan, Guangdong 523808, China}

\date{\today}

\begin{abstract}
    The interplay of topological band structures and electronic correlations may lead to novel quantum phenomena with potential applications. First-principles calculations are critical for guiding experimental discoveries and interpretations, but often fail if electronic correlations cannot be properly treated. Here we show that this issue occurs also in the kagome altermagnets Mn$_3X$ ($X=$ Sn, Ge, Ga), which were believed to exhibit large anomalous Hall effect due to topological band structures with Weyl nodes near the Fermi energy. Our systematic investigations reveal critical importance of beyond-DFT treatments on three key aspects of their magnetic, electronic, and topological properties: (1) establishment of noncollinear altermagnetic orders, (2) weakly renormalized band structures in excellent agreement with angle-resolved photoemission spectroscopy experiment, and (3) sensitive tuning of the Weyl nodes. Our work provides a unified basis for understanding topological properties of the Mn$_3X$ family, which challenges previous experimental interpretations based on DFT band structures and predicts potentially higher anomalous Hall conductivity in Mn$_3$Ga under electron doping. This  underscores the importance of a correlation-aware framework beyond DFT in understanding topological magnetic materials.
\end{abstract}

\maketitle

Topological quantum materials have attracted intensive interest for their importance on both fundamental physics and potential applications. Increasing attentions have recently focused on  correlated topological materials, in which exotic many-body phenomena may emerge due to the interplay of nontrivial topology and electronic correlations \cite{Dzero2016ARCMP, Chen2022NP}. One notable example is Mn$_3$Sn, which adopts a kagome bilayer structure and exhibits a noncollinear inverse triangular antiferromagnetic (AFM) order in the kagome plane below 430 K\cite{Brown1990JPCM,Nakatsuji2015N}. This compound has a considerable spin-orbit coupling (SOC) that breaks the effective time reversal symmetry and may cause ferromagnetic behaviors such as nonzero Berry curvatures and sizable anomalous Hall conductivities without external magnetic field \cite{Suzuki2017PRB}. A large anomalous Hall effect (AHE) has indeed been observed at room temperature in spite of the vanishingly small net magnetization \cite{Nakatsuji2015N}. Resistivity and thermal conductivity measurements reported an anomalous Lorenz number and excluded its origin from inelastic scattering\cite{Li2017PRL}. A large magneto-optical Kerr effect (MOKE) was also observed, implying the presence of analogous symmetry requirement for the AHE \cite{Higo2018NP}. Hence, Mn$_3$Sn has been established to be the first AFM material with intrinsic AHE. Similar exotic properties have later been reported in the isostructural compound Mn$_3$Ge \cite{Nayak2016SAa,Kiyohara2016PRAa, Wuttke2019PRB, Wu2020APL}. Recently, an even larger AHE was observed in the Ga-rich Mn$_3$Ga \cite{Song2024AFM}. Understanding the origin of these exotic properties is therefore highly desired.

First-principles calculations of electronic band structures played a key role in guiding the above experimental discoveries and interpretations. Density functional theory (DFT) studies first predicted the existence of Weyl points near the Fermi energy \cite{Kubler2014EPL,Yang2017NJP,Kubler2017E}, which are singular points of Berry curvatures (magnetic monopoles in momentum space) \cite{Fang2003S,Ma2021NC} and may give rise to a large AHE. Comparison with the angle-resolved photoemission spectroscopy (ARPES) experiment in Mn$_3$Sn \cite{Kuroda2017NM} seems to support the predicted DFT band structures, but requires a very large band renormalization factor of about $5$, suggesting significant electronic correlations. In sharp contrast, comparison between DFT and ARPES measurements on Mn$_3$Ge \cite{Changdar2023} only requires a small renormalization factor of 1.18. This large discrepancy between two isostructural compounds with similar exotic properties also calls for reinvestigation of their correlated electronic structures, which typically cannot be captured by DFT. In addition, the Hund's rule coupling also cannot be well treated by DFT but has been shown to be important for the partially filled $3d$ shell \cite{Cao2024PRB, Mravlje2011PRL}, especially in Mn-based compounds including Mn$_3$Sn \cite{Georges2013ARCMP,Yu2022PRB}. This raises the concern on the reliability of the DFT-predicted band structures. The issue becomes particularly important considering that DFT predictions of topological band structures with the Weyl points have been widely adopted in experimental interpretations, although their presence has not been visualized in ARPES. Recently, DFT+U calculations have even been applied to systematically identify possible topological phase transition with the onsite Coulomb interaction \cite{Xu2020N}. 

In this work, we show that the aforementioned controversy can be resolved in a unified manner through correlated electronic calculations beyond DFT, and a consistent description is achieved across the whole Mn$_3X$ family by combining DFT \cite{2014WIEN2k} and dynamical mean-field theory (DMFT) \cite{Georges1996RMP, Anisimov1997JPCM, Lichtenstein1998PRB, Kotliar2006RMP, Held2008JPCM, Haule2010PRB} with full inclusion of electron interactions and spin-orbit coupling. Our calculations successfully reproduce ARPES spectra of Mn$_3$Sn and Mn$_3$Ge with remarkable details, confirming a universal band renormalization factor ($Z^{-1} \approx 1.6$) that reconciles previous discrepancies among DFT-based estimates. Our theory further captures the evolution of topological node structures, including the potential formation of line nodes in Mn$_3$Sn, unattainable in standard DFT treatments. The calculated anomalous Hall conductivities (AHC) are in qualitative consistency with experiments for the whole series, which provides a coherent basis for understanding their topological properties. We further predict that even higher AHC may be realized by electron doping in Mn$_3$Ga. These demonstrate how a systematic beyond-DFT framework incorporating key interactions can yield a unified and material-specific understanding of complex magnetic and topological orders.\\

\begin{figure}
    \begin{center}
        \includegraphics[width=0.48\textwidth]{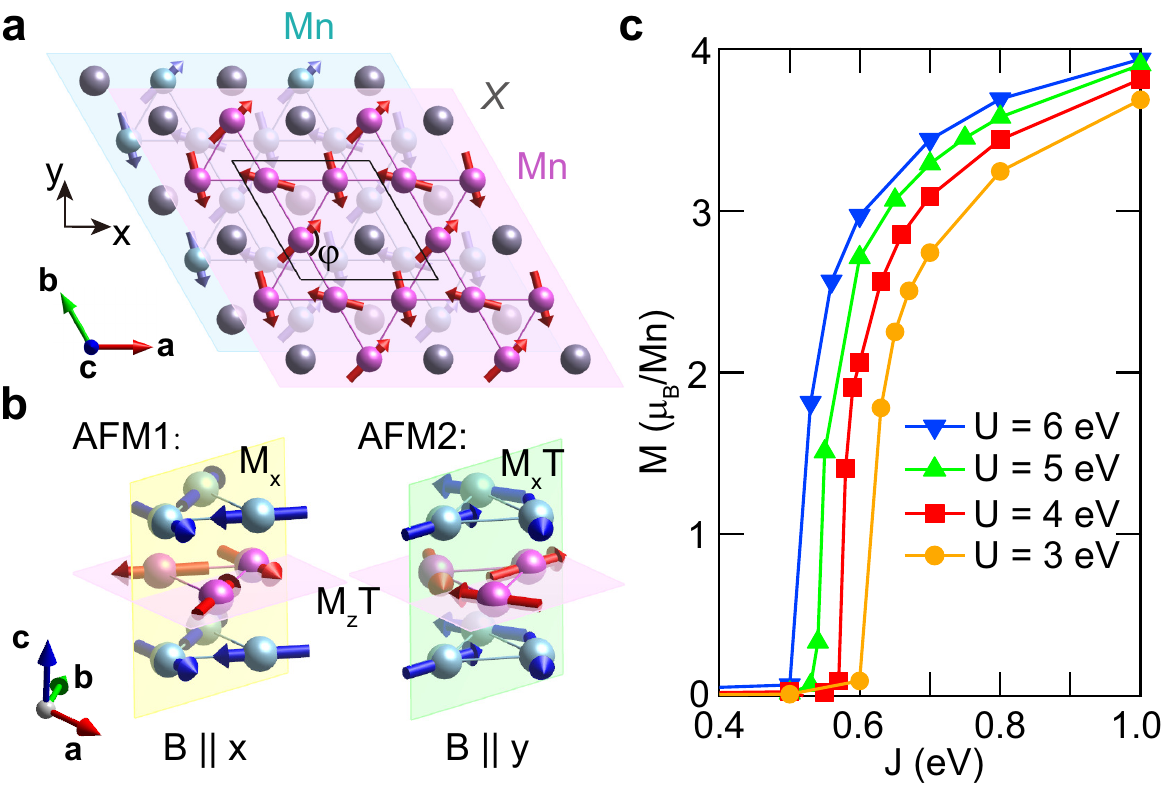}
        \caption{\textbf{Crystal and magnetic properties.} \textbf{a} Illustration of the crystal structure of Mn$_3X$ with inverse triangular AFM configurations showing inverse symmetry between two adjacent layers \cite{VESTA}. \textbf{b} Symmetry of the inverse triangular AFM order for two spin configurations, AFM1 and AFM2. Three planes are shown for $M_zT$ (magenta), $M_x$ (yellow), and $M_xT$ (green), where $M_x$ and $M_z$ are mirror operations and $T$ is the time reversal operation. \textbf{c} The spin moment of a single Mn-ion in Mn$_3$Sn for different choices of $U$ and $J$ calculated using DFT+DMFT at 300~K.
        }\label{fig1}
    \end{center}
\end{figure}

\noindent\textbf{Results}

\noindent\textbf{Structural and magnetic analyses.}
Mn$_3X$ adopt a hexagonal structure with the space group $P6_3/mmc$. The lattice constants are $a=5.67\,$\AA{}, $c=4.53\,$\AA{} for Mn$_3$Sn \cite{Brown1990JPCM},  $a=5.338\,$\AA{}, $c=4.312\,$\AA{} for Mn$_3$Ge \cite{Gupta1970JotLCM}, and $a=5.404\,$\AA{}, $c=4.357\,$\AA{} for Mn$_3$Ga \cite{Niida1983JPSJ}. The Mn-ions form a kagome lattice with mixed triangles and hexagons in the $ab$-plane, and the $X$-ions are located at the center of the hexagons. As shown in Fig.~\ref{fig1}a, the adjacent layers stack along the $c$-axis with an offset. Below the N\'eel temperature, a noncollinear AFM state is formed with Mn moments aligned in an inverse triangular spin structure within the $ab$-plane \cite{Brown1990JPCM, Soh2020PRB,Kren1970SSC}. As the temperature decreases below $200$~K, a helical or incommensurate spin structure may appear in Mn$_3$Sn depending sensitively on the samples and suppress the AFM until a spin glass is observed below about $50$~K \cite{Sung2018APL}. For simplicity, we only consider the AFM phase, in which the inverse symmetry is preserved to transform the two kagome layers to each other. Figure~\ref{fig1}b shows two quenched spin configurations of the inverse triangular AFM with higher symmetry, AFM1 and AFM2. Mn$_3$Ge is reported to adopt AFM2 order as the ground state \cite{Soh2020PRB}, while Mn$_3$Sn switches between the two depending on the direction of the applied magnetic field \cite{Tomiyoshi1982JPSJ}. In contrast, Mn$_3$Ga takes a lower-symmetry configuration as shown in Fig.~\ref{fig1}a with $\varphi = 105^{\circ}$\cite{Kren1970SSC}. According to the classification in Ref. \onlinecite{Cheong2024nQM}, they have the ferromagnetic point group and belong to type-I altermagnets with alternating spin orientations and local environments simultaneously, which should exhibit linear AHE with non-zero net magnetic moments. All three configurations satisfy the inverse symmetry and the combined symmetry $M_{z}T$, and the coplanar AFM1 (AFM2) order has an additional mirror symmetry $M_{x}$ ($M_{x}T$). 
While the inverse symmetry does not affect the Berry curvatures, the mirror reflection reverses the components of the Berry curvatures parallel to the mirror plane and the time reversal symmetry ($T$) reverses the Berry curvatures along all directions. As a result, the Hall conductivity given by integrating the Berry curvatures over the whole Brillouin zone is nonzero only along the $x$ direction for AFM1 and along the $y$ direction for AFM2, but nonzero along both directions for the lower-symmetry configuration. In real materials, $M_{x}$ and $M_{x}T$ may be slightly broken due to a tiny net moment in $ab$-plane \cite{Nayak2016SAa}, causing nonzero AHC along both $x$ and $y$ directions. This is often ignored in first-principles calculations for simplicity \cite{Kubler2017E,Kuroda2017NM}. Our DFT+DMFT results in Supplementary Fig.~S3 confirm that it is indeed a small perturbation to the band structures. We will therefore focus on the ideal inverse triangular AFM orders here.\\

\noindent\textbf{Stabilization of magnetic order.}
We first consider Mn$_3$Sn with the AFM1 order, whose electronic band structures have been measured by ARPES \cite{Kuroda2017NM}. Figure~\ref{fig1}c plots the calculated Mn spin moment as functions of the Hubbard $U$ and the Hund's rule coupling $J$, whose values have been chosen following the literature \cite{Chen2020NC, Li2021NC}. The calculations were carried out using DFT+DMFT at 300~K below $T_{\text N}=430\,$K, showing that the magnetic order can only be stable for sufficiently large $U$ and $J$. The critical $J_{\text c}$ lies roughly between 0.5 and 0.6 eV and varies slightly when $U$ increases from 3 to 6 eV. Its increase with decreasing $U$ indicates their compensating role for local spin polarization. Without the Hund's rule coupling, the moment no longer exists regardless of the value of $U$, at least within our calculated parameter range. By contrast, the size of the moment grows quickly as $J>J_{\text c}$, implying its sensitive dependence on the Hund's rule coupling, as was also shown in previous study \cite{Yu2022PRB}. We thus conclude that the Hund's rule coupling plays a key role in establishing the altermagnetism in Mn$_3$Sn.

\begin{figure}
    \begin{center}
        \includegraphics[width=0.48\textwidth]{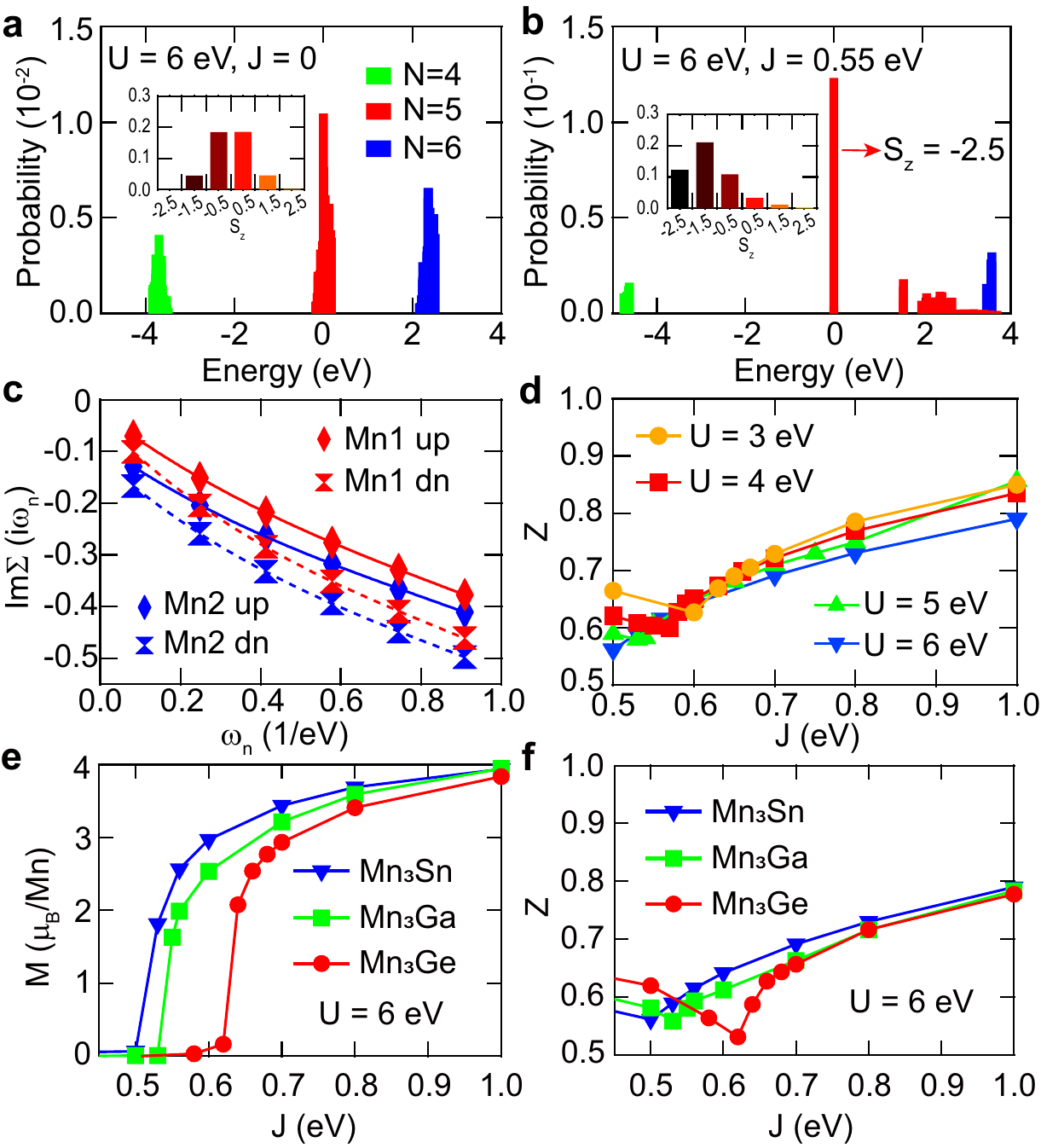}
        \caption{\textbf{Effect of the Hund's rule coupling on the Mn-ionic states and band renormalization.} \textbf{a, b} Probabilistic distribution of the DMFT state of Mn$_3$Sn on different Mn-ionic configurations with the electron occupancy $N=4$, 5, 6 with ($J=0.55\,$eV) and without the Hund's rule coupling at $U =6\,$eV. The horizontal axis denotes their respective energy difference with respect to the configuration of largest probability. The insets show the distribution of $N=5$ state on the local $z$-component of the total spin, $S_z$. \textbf{c} Illustration of the fourth-order polynomial fitting (solid and dashed lines) to the imaginary part of the orbital-averaged self-energy as a function of the Matsubara frequency for $U=6\,$eV and $J=0.55\,$eV for Mn$_3$Sn. \textbf{d} The renormalization factor $Z$ estimated from the DFT+DMFT self-energies with varying $U$ and $J$ for Mn$_3$Sn. \textbf{e} Comparison of the spin moment per Mn-ion in Mn$_3$Sn, Mn$_3$Ge, and Mn$_3$Ga as functions of the Hund's rule coupling $J$ calculated using DFT+DMFT for a fixed $U=6\,$eV at 300~K. \textbf{f} Comparison of their renormalization factor $Z$ extracted from the DFT+DMFT self-energies with varying $J$ for $U=6\,$eV at 300~K.}\label{fig2}
    \end{center}
\end{figure}

To understand what happens, we plot in Figs.~\ref{fig2}a and \ref{fig2}b the projection of the calculated DMFT states on the ionic configurations $(N,S_z)$ of the Mn-$3d$ orbitals for a fixed $U=6\,$eV and different values of $J$ \cite{Shim2007N}. Here $N$ is the occupation number of electrons on Mn-$3d$ orbitals and $S_z$ is the local $z$-component of the total spin. For $J=0$ shown in Fig.~\ref{fig2}a, we find strong fluctuations among the three valence states $N=4$, 5, 6 indicated by the similar probabilities of DMFT state on these configurations. Meanwhile, the spin also distributes almost equally along opposite directions, with the highest probability in the $S_z=\pm 0.5$ low spin states. The system is in a nonmagnetic phase despite of the large Hubbard $U=6\,$eV. By contrast, once the Hund's rule coupling $J=0.55\,$eV is switched on as shown in Fig.~\ref{fig2}b, the $N=5$, $S_{z}=-2.5$ state becomes dominant and the Mn-spins become polarized, with most of the weight on negative $S_z$ fluctuating strongly among $S_z=-2.5$, -1.5, and -0.5. The total spin moment is thus about 2.6 $\mu_{\text B}$/Mn, which is smaller than the DFT prediction of 3.3 $\mu_{\text B}$ but consistent with experiments \cite{Brown1990JPCM,Tomiyoshi1982JPSJ} where the estimated value of magnetic moment varies from 1.78 to 3.0 $\mu_B$ depending on the detection technologies, temperature and samples. We will show that this difference is crucial for yielding a spectral function consistent with ARPES measurement \cite{Kuroda2017NM}.\\

\noindent\textbf{Band renormalization.}
The strength of electronic correlations may be seen straightforwardly from the band renormalization. Figure~\ref{fig2}c plots the imaginary part of the orbital-averaged self-energies $\Sigma(i\omega)$ as a function of Matsubara frequency for different spin directions and inequivalent Mn-ions. We find that they behave similarly on two inequivalent ions, possibly due to their similar crystal field environments. The solid and dashed lines give the fourth-order polynomial fitting, which is used to estimate the renormalization factor $Z^{-1}=1-\partial\text{Im}\Sigma(i\omega)/\partial\omega|_{\omega\to 0^{+}}$ \cite{Mravlje2011PRL}. For clarity, we take the average over all Mn-$3d$ orbitals in a unit cell and plot the derived $Z$ as a function of $U$ and $J$ in Fig.~\ref{fig2}d. Interestingly, $Z$ increases gradually with increasing $J$ in the AFM phase for $J>J_c$, but behaves oppositely in the paramagnetic phase. This trend may come from the enhanced spin fluctuations near the critical point \cite{Schroder1998PRL}. At $J_c$, we obtain a maximum band renormalization $Z^{-1}$, which is, however, less than 1.8 and hence three times smaller than previous estimate from the comparison between DFT and ARPES in Mn$_3$Sn \cite{Kuroda2017NM}. 

The behavior observed in Mn$_3$Sn is consistently reproduced across the Mn$_3X$ family. Figure~\ref{fig2}e displays the Mn spin moments for Mn$_3$Ge and Mn$_3$Ga under their respective in-plane AFM orders, obtained from DFT+DMFT calculations with a fixed $U = 6,$eV, and Fig.~\ref{fig2}f shows the corresponding quasiparticle renormalization factors $Z$. Again, there exists a critical Hund's rule coupling $J_c$, which increases to about $0.55\,$eV for Mn$_3$Ga and $0.62\,$eV for Mn$_3$Ge, above which the spin moment first grows rapidly and then gradually saturates to about 4 $\mu_{\text{B}}$/Mn at $J=1\,$eV. We have tested both inverse AFM orders for Mn$_3$Ge and found almost the same $J_c$. The variation of $J_c$ from Mn$_3$Sn to Mn$_3$Ga and Mn$_3$Ge is in conformity with their lattice parameters, indicating a competition between the kinetic energy and the Hund correlation. In all cases, the renormalization factor $Z$ is the smallest at the transition and varies between 0.5 and 0.8 for $J$ up to 1 eV. We have therefore a unified scenario for the band renormalization and moment formation in the whole Mn$_3X$ family, in contrast to previous DFT-ARPES speculations \cite{Kuroda2017NM,Changdar2023,Chen2021NC}.\\

\begin{figure}[t]
    \begin{center}
        \includegraphics[width=0.48\textwidth]{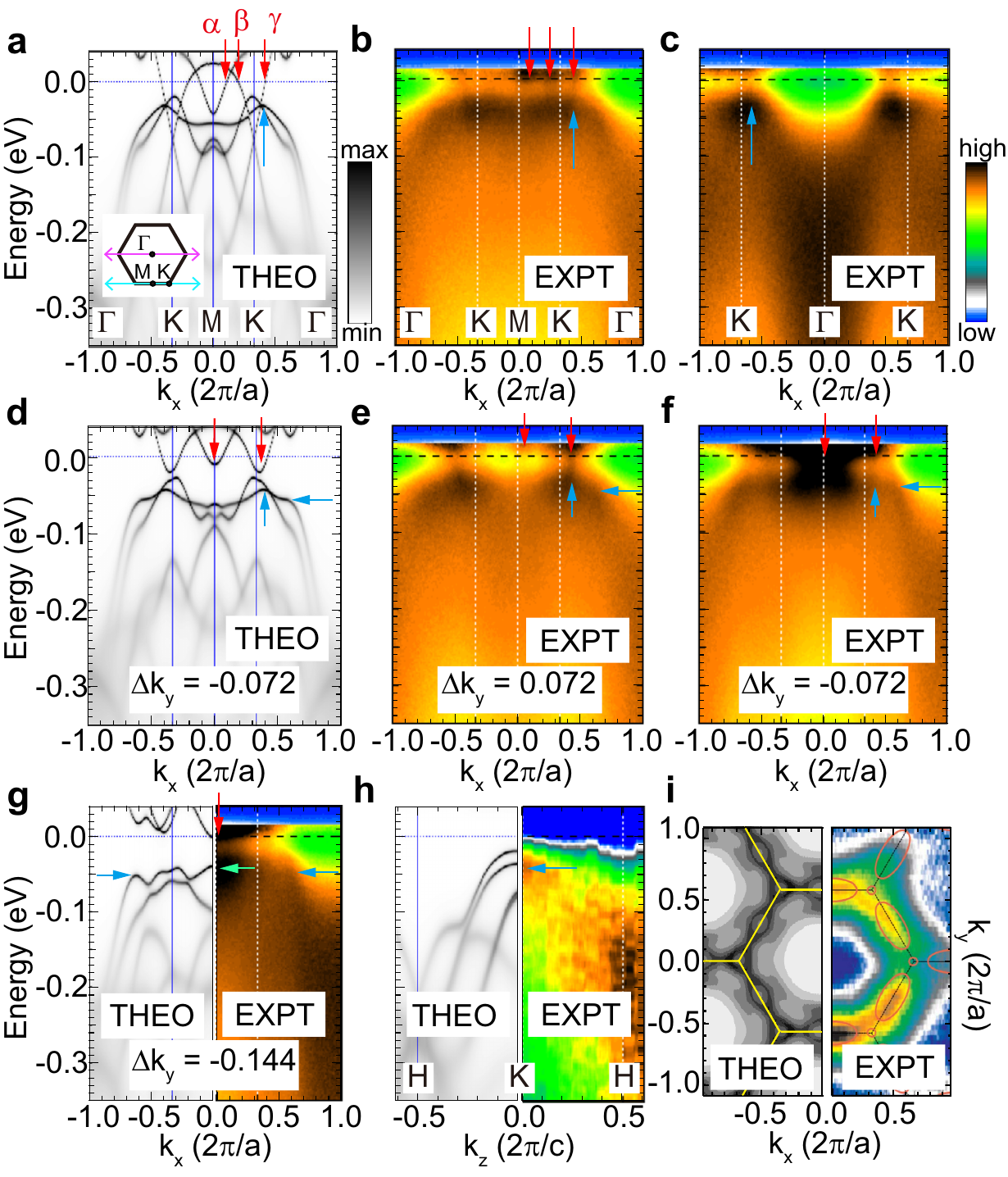}
        \caption{\textbf{Comparison of ARPES \cite{Kuroda2017NM} and DFT+DMFT band structures in Mn$_3$Sn.} Intensity map of the theoretical and experimental spectral functions along: \textbf{a, b} the high-symmetry $\Gamma$-K-M-K-$\Gamma$ line (cyan arrows in the inset of \textbf{a}); \textbf{c} the high-symmetry M-K-$\Gamma$-K-M  line (magenta arrows in the inset of \textbf{a}); \textbf{d}-\textbf{f} the $k_x$ direction with $|\Delta k_{y}|=0.072\,(2\pi/a)$ slightly off the $\Gamma$-K-M-K-$\Gamma$ path; \textbf{g} the $k_x$ direction with $\Delta k_{y}=-0.144\,(2\pi/a)$ from $\Gamma$-K-M-K-$\Gamma$; \textbf{h} the high-symmetry H-K-H line. The arrows of different colors highlight some characteristic features of excellent agreement. \textbf{i} Comparison of the theoretical and experimental Fermi surface mapping in the $k_z=0$ plane. To mimic the low resolution of the ARPES data, we have intentionally chosen the DFT+DMFT results at 300~K. The calculated Fermi surfaces are qualitative the same at 60~K. The yellow line in the left one indicates the hexagonal Brillouin zone. The red curves show the DFT-predicted Fermi surfaces reproduced from Ref. \citenum{Kuroda2017NM}.}\label{fig3}
    \end{center}
\end{figure}

\noindent\textbf{Comparison with ARPES on Mn$_3$Sn.}
The much smaller renormalization factor for Mn$_3$Sn questions the validity concerning the interpretation of its ARPES data using the artificially renormalized DFT band structures. In fact, as shown in Supplementary Fig.~S1, the renormalized DFT bands fail to conform to experimental observations in many aspects even on a qualitative level. Surprisingly, we find that DFT+DMFT can actually give an excellent agreement with the ARPES data for reasonable parameters. Figure~\ref{fig3} shows the intensity map of the spectral function along different line cuts in the Brillouin zone. The ARPES data were reproduced from Ref. \onlinecite{Kuroda2017NM} and the calculations were performed for $U=6\,$eV and $J=0.55\,$eV at $T=60\,$K. As compared in Figs.~\ref{fig3}a-\ref{fig3}c, DFT+DMFT successfully captures several major features of the ARPES data along the high-symmetry path M-K-$\Gamma$, where three bands labeled as $\alpha, \beta$, and $\gamma$ cross the Fermi energy (red arrows) and a flat band appears around M at about -0.05~eV. The latter exhibits a peak near the K point (blue arrows), then bends down, and reaches about -0.3~eV at $\Gamma$. By contrast, DFT gives a flat band close to the M point at about -0.25~eV (Supplementary Fig.~S1), which is far away from the Fermi energy and hence has to be renormalized by a factor of 5 to explain the ARPES observation. However, such a large artificial renormalization introduces significant spectral weights between -0.3 and -0.1 eV near the K-M path (Supplementary Fig.~S1), a feature absent in experiment \cite{Kuroda2017NM}.

Away from the high-symmetry path by $\Delta k_y =-0.072$ (in unit of 2$\pi/a$), two bands (red arrows) were observed to cross the Fermi energy along $k_x$ in ARPES, as plotted in Figs.~\ref{fig3}e and \ref{fig3}f. This feature is well captured in DFT+DMFT as marked by the arrows in Fig.~\ref{fig3}d. The numerical results provide a better view on the band evolution with $k_y$. Further away from the high-symmetry path, the conduction bands gradually move upwards and the number of bands crossing the Fermi energy decreases. For $\Delta k_y =-0.144$ in Fig.~\ref{fig3}g, only the $\alpha$ band (red arrow) crosses the Fermi energy. The bands that intersect around K start to open a gap, display a shoulder (blue arrows),  and bulge upward at $k_x=0$ (green arrow). All features are in good consistency with experiment. 

The dispersion along another high-symmetry line K-H is also compared in Fig.~\ref{fig3}h. The quantitative agreement with ARPES again supports our DFT+DMFT calculations, while in DFT, the renormalization would cause an almost flat band along K-H as shown in Supplementary Fig.~S1, which is missing in ARPES. Figure~\ref{fig3}i compares the projected Fermi surfaces on $k_z=0$ plane. We find a better agreement between ARPES and DFT+DMFT. In particular, both have larger Fermi surfaces around some K points than those of DFT (red lines).

These band features depend sensitively on the Hund's rule coupling as discussed in detail in Supplementary Sec. B and Fig.~S2. For a slightly larger $J=0.6\,$eV, the flat band around M moves further away from the Fermi energy to about -0.1 eV, and many other features also deviate from ARPES observations. This highlights the importance of the Hund's rule coupling via tuning the spin states of the Mn-ions, and explains why DFT fails to yield the correct band structures even with a large renormalization factor.\\

\begin{figure}
    \begin{center}
        \includegraphics[width=0.48\textwidth]{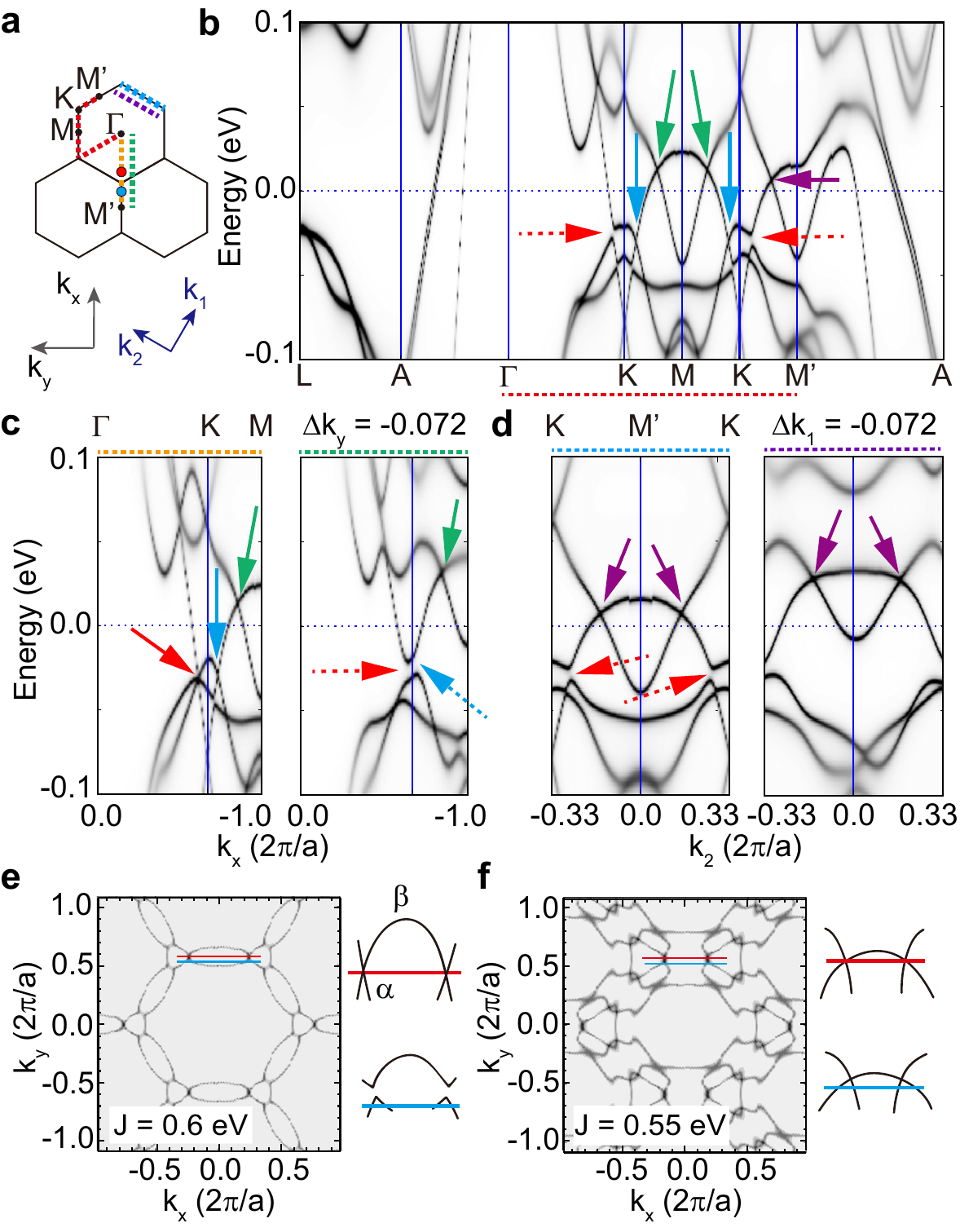}
        \caption{\textbf{Effect of the Hund's rule coupling on the Weyl points in Mn$_3$Sn.} \textbf{a} Illustration of the selected $\mathbf{k}$-paths marked by dashed lines of different colors on the $k_z=0$ plane of the Brillouin zone. \textbf{b} The spectral  function along the high-symmetry line L-A-$\Gamma$-K-M-K-M$'$-A, where $\Gamma$-K-M-K-M$'$ is denoted by the red dashed line in \textbf{a}. The solid (dashed) arrow marks the band crossing (gap). \textbf{c} Comparison of the spectral functions along $\Gamma$-K-M (orange dashed line) and the line shifted by $\Delta k_{y}=-0.072\,(2\pi/a)$ (green dashed line). Note that the orange $\Gamma$-K-M line is different from the red one because of the lack of six-fold rotational symmetry. \textbf{d} Comparison of the spectral function along K-M$'$-K (blue dashed line) and the line shifted by $\Delta k_{1}=-0.072\,(2\pi/a)$ (purple dashed line). \textbf{e, f} The band maps on the $k_z=0$ plane at $E=9\,$meV for $J=0.6\,$eV and $E=50\,$meV for $J=0.55\,$eV, illustrating the evolution of the band crossings with the Hund's rule coupling.}\label{fig4}
    \end{center}
\end{figure}

\noindent\textbf{Tuning the Weyl points.}
Weyl points in Mn$_3X$ were predicted to be present near the Fermi energy in DFT and may contribute significantly to the AHE \cite{Kubler2014EPL, Yang2017NJP, Kubler2017E}. Their presence in Mn$_3$Sn was supported by the observation of a large anomalous Nernst effect beyond that in ferromagnetic metals \cite{Ikhlas2017NP} and the chiral anomaly in magnetotransport\cite{Chen2021NC}. To identify the Weyl points, we study in detail the DFT+DMFT band structures for $U=6\,$eV and $J=0.55\,$eV along different paths in the Brillouin zone marked by dashed lines of different colors in Fig.~\ref{fig4}a. Figure~\ref{fig4}b shows several crossing points of non-degenerate linear bands along the high-symmetry path L-A-$\Gamma$-K-M-K-M$'$-A (red dashed line): two at about -30 meV between K and M marked by blue arrows, two at about 13 meV marked by green arrows, and one at about 8 meV marked by the purple arrow. In addition, there are two small gaps (red arrows) near K along the $\Gamma$-K and K-M$'$ paths. We first focus on the dispersion near K and M and take the path along the $k_x$ direction where the existence of Weyl points was predicted above the Fermi energy in DFT calculation \cite{Kuroda2017NM}. In Fig.~\ref{fig4}c, it is seen that among all three band crossings along $\Gamma$-K-M (orange dashed line), only those around the K point (red and blue arrows) become gapped when the path is moved slightly by $\Delta k_y=-0.072$, while the one around M (green arrows) remains unchanged. These indicate that those marked by red and blue arrows are Weyl points, while the one marked by green arrow is a part of line nodes. Note that the $\Gamma$-K path (orange dashed line) in Fig.~\ref{fig4}c is different from the $\Gamma$-K path (red dashed line) in Fig.~\ref{fig4}b. Along the K-M$'$-K path in Fig.~\ref{fig4}d, the two crossings marked by purple arrows remain unchanged away from the high-symmetry path and are also parts of line nodes. For clarity, we mark the positions of the Weyl points by red and blue dots in Fig.~\ref{fig4}a. Both are located around K and at about -30 meV below the Fermi energy, in contrast to the DFT-predicted Weyl points above the Fermi energy. In addition, the line nodes obtained here above the Fermi energy are not present in previous DFT calculations. 

To understand the origin of these differences, we compare the band maps on the $k_z=0$ plane for $J=0.6\,$ and 0.55 eV. For $J=0.6\,$eV shown in Fig.~\ref{fig4}e, two bands cross at 9~meV along K-M-K (red line) but become gapped slightly off this path (blue line). They correspond to the Weyl points around K in DFT calculations and exist here for $J$ beyond 0.6~eV. As $J$ decreases, the top of the $\beta$ band and the bottom of the $\alpha$ band are renormalized towards the Fermi energy, and their dispersions start to deviate from linearity near the nodes. For $J=0.55\,$eV, these nodes are pushed to about 50~meV but remain unaffected under a slight shift from the high-symmetry path, so they are no longer Weyl points (Fig.~\ref{fig4}f). In the meanwhile, the bands around K point at -0.1~eV that bend down at $J=0.6\,$eV are lifted towards the Fermi energy and renormalized to bend upwards at $J=0.55\,$eV. They eventually intersect with the $\beta$ and $\gamma$ bands around K and produce the Weyl points below the Fermi energy in Fig.~\ref{fig4}c. Thus, in Mn$_3$Sn, these Weyl points start to show up only below $J\approx 0.6\,$eV and move closer to the Fermi energy as $J$ further decreases. They are not present in prior first-principles calculations as shown in Supplementary Fig.~S1. Their sensitive dependence on the Hund's rule coupling may be attributed to the rapid change of the Mn spin moment as $J$ exceeds its critical value $J_c$, which in turn changes the band dispersion as well as the relative band shift of different spin components (Supplementary Fig.~S2).\\

\noindent\textbf{Mn$_3$Ge and Mn$_3$Ga.}
\begin{figure}
    \begin{center}
        \includegraphics[width=0.48\textwidth]{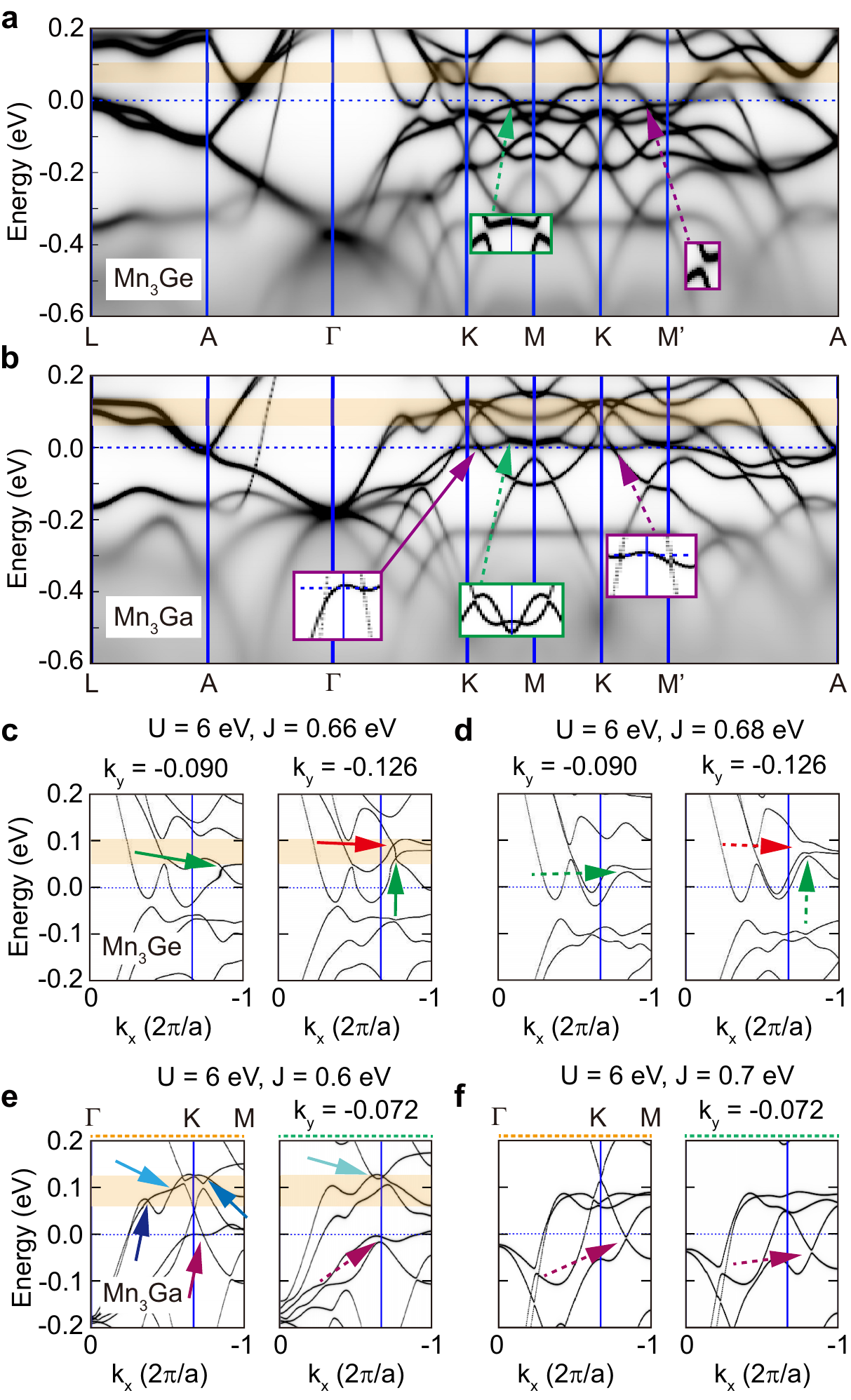}
        \caption{\textbf{DFT+DMFT electronic band structures of Mn$_3$Ge and Mn$_3$Ga.} \textbf{a, b} The DFT+DMFT spectral function along the high-symmetry line L-A-$\Gamma$-K-M-K-M$'$-A for Mn$_3$Ge with the AFM2 order at $J=0.66\,$eV, $U=6\,$eV, $T=300\,$K and Mn$_3$Ga with the lower-symmetry configuration at $J=0.6\,$eV, $U=6\,$eV, $T=300\,$K. \textbf{c} Band crossings of Mn$_3$Ge along the path shifted by $\Delta k_y = -0.090$ and -0.126 (2$\pi$/a) away from $\Gamma$-K-M (orange dashed line in Fig.~\ref{fig4}a) for $U=6\,$eV and $J = 0.66\,$eV. \textbf{d} Gap opening at $J=0.68\,$eV in Mn$_3$Ge, showing sensitive tuning by the Hund's rule coupling. \textbf{e, f} Comparison of the band structures of Mn$_3$Ga along $\Gamma$-K-M and that shifted by $\Delta k_y = -0.072$ (2$\pi$/a) (green dashed line in Fig.~\ref{fig4}a) 
        for $U=6\,$eV and $J = 0.6$ and 0.7~eV, respectively. The solid arrows mark the band crossing and the dashed arrows mark the gap.}
        \label{fig5}
    \end{center}
\end{figure}
Having established some basic facts for Mn$_3$Sn, we now extend the discussions to other members of the Mn$_3X$ family. The DFT+DMFT spectral function for Mn$_3$Ge is plotted in Fig.~\ref{fig5}a along the same high-symmetry path. We use $J=0.66\,$eV so that the calculated spin moment is close to the experimental value of 2.65~$\mu_{\text{B}}$/Mn in Mn$_3$Ge \cite{Soh2020PRB}. Again, DFT+DMFT calculations result in a small $Z^{-1}\approx 1.6$ as in Mn$_3$Sn. Flat bands are clearly seen around -0.05 eV along the M-K path and bend steeply downwards halfway along K-$\Gamma$. Around $\Gamma$, there are additional dispersive bands bending down towards K and A below -0.38 eV. All these features are in good consistency with latest ARPES measurements \cite{Changdar2023} without artificial renormalization. The overall consistencies for both Mn$_3$Ge and Mn$_3$Sn validate our systematic DFT+DMFT calculations. For comparison, Fig.~\ref{fig5}b shows the spectral function of Mn$_3$Ga for an intermediate $J=0.6\,$eV, which has yet to be measured in experiment but looks similar to Mn$_3$Sn and Mn$_3$Ge in many aspects, except that the flat bands along M-K now shift to about 0.1 eV above the Fermi energy because Ga atoms contribute less valence electrons. 

However, a closer examination reveals important differences that are crucial for their large AHE. For Mn$_3$Ge, as shown in Fig.~\ref{fig5}a, no band crossing is observed around the Fermi energy along the high-symmetry path at $J=0.66\,$eV. In fact, Weyl points are prohibited along $\Gamma$-K-M in Mn$_3$Ge, because the mirror and inversion symmetry would reverse the chirality of any potential Weyl nodes. Instead, we find band crossings along the paths shifted by $\Delta k_y=-0.090$ and -0.126 separated by the gap at around $\Delta k_y=-0.108$ away from the high-symmetry path, see more details in Fig.~\ref{fig5}c and Supplementary Fig.~S4. These Weyl points are located from 0.04 to 0.09 eV above the Fermi energy, suggesting enhanced AHC with electron doping. Again, they are very sensitive to the value of $J$, as compared for $J=0.68\,$eV in Fig.~\ref{fig5}d.

By contrast, the Weyl points can be clearly identified near the Fermi energy in Mn$_3$Ga for $J=0.6\,$eV, as marked by the purple solid arrows illustrated in the inset of Fig.~\ref{fig5}b and demonstrated by the gap opening away from the high-symmetry path $\Gamma$-K-M by $\Delta k_y= -0.072$ in Fig.~\ref{fig5}e. We also identify two small gaps from the band bending near the M point and along the K-M$'$ path. Upon electron doping, we observe multiple Weyl points along the high-symmetry path between 0.07 to 0.12 eV above the Fermi energy. Unlike in Mn$_3$Ge, they are no longer prohibited due to the lower-symmetry magnetic order in Mn$_3$Ga. As the Hund's rule coupling $J$ increases to 0.7 eV, the Weyl point at Fermi level turns into a small gap as shown in Fig.~\ref{fig5}f. These results demonstrate that the topological properties across the Mn$_3X$ family are highly sensitive to the strength of electronic correlations and the resulting magnetic orders. As a consequence,  the Weyl points are accidental and not symmetry-protected \cite{Xu2020N}.\\

\noindent\textbf{Discussion and conclusions}\\
\begin{figure}
    \begin{center}
        \includegraphics[width=0.48\textwidth]{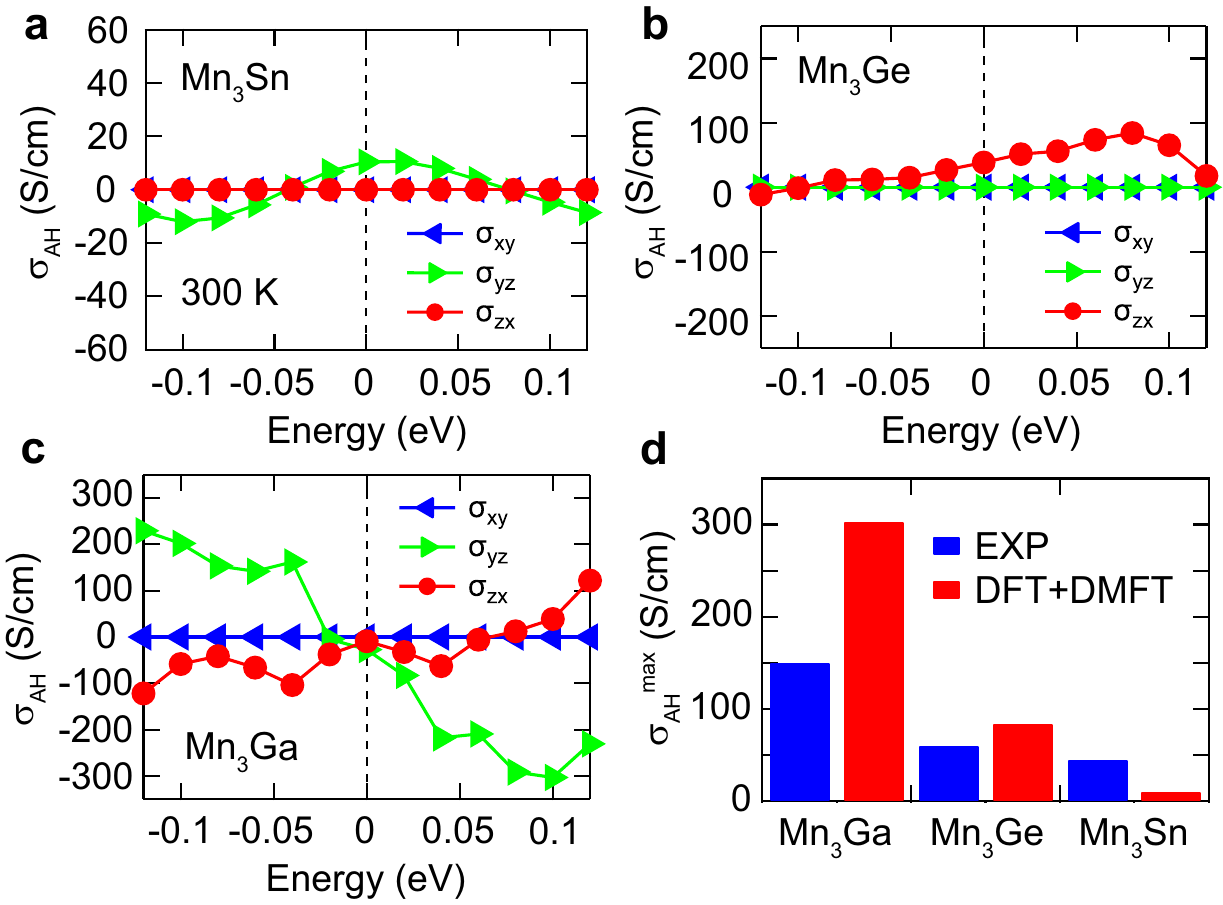}
        \caption{\textbf{Variation of the AHC in Mn$_3X$.} \textbf{a}-\textbf{c} Calculated AHC by DFT+DMFT as a function of doping represented by the Fermi energy shift for Mn$_3$Sn, Mn$_3$Ge, and Mn$_3$Ga at 300~K. \textbf{d} Comparison of the maximum magnitude of AHC at 300~K from DFT+DMFT calculations (in red) and experimental values (in blue) \cite{Kuroda2017NM,Chen2021NC,Song2024AFM}.}\label{fig6}
    \end{center}
\end{figure}
The subtle differences between DFT and DFT+DMFT band structures may have a significant impact on experimental interpretations. Currently, direct experimental evidences are lacking for precise locations of the Weyl points, so most data analyses have relied heavily on DFT band structures. It has been suggested that increasing Mn content in Mn$_3$Sn and Mn$_3$Ge could introduce additional Mn-$d$ electrons, move the Fermi energy upwards towards the DFT-predicted Weyl points, and thereby enhance the AHC \cite{Kuroda2017NM,Chen2021NC}. However, in the DFT+DMFT band structures of Mn$_3$Sn, the Weyl points are located below the Fermi energy for appropriate parameters, which questions the role of the Weyl points. To resolve this contradiction, we further calculate the AHC as a function of doping represented by the Fermi energy shift. The results are shown in Figs.~\ref{fig6}a for Mn$_3$Sn and \ref{fig6}b for Mn$_3$Ge. Because of symmetry, the only nonzero component is along the $x$ direction for Mn$_3$Sn (AFM1) and the $y$ direction for Mn$_3$Ge (AFM2). We still find that both are enhanced as the Fermi energy moves slightly upwards. But for Mn$_3$Sn, the AHC only increases slightly and reaches the maximum when the Fermi energy is shifted upwards by about 0.02 eV, which is close to the position of the nodal lines. This suggests some contributions from the nodal line structure, which has also been discussed previously in Co$_2$MnAl \cite{Li2020NC}, although the major increase seems to occur above -0.1 eV from the Weyl points below the Fermi energy. On the other hand, for Mn$_3$Ge shown in Fig.~\ref{fig6}b, the maximum AHC is achieved at about 0.08 eV, where we find Weyl points in the DFT+DMFT spectral function (Fig.~\ref{fig5}c). Figure~\ref{fig6}c shows the calculated AHC along both $x$ and $y$ directions for Mn$_3$Ga. We see sensitive dependence of the AHC on electron doping, where Ga-rich is fitted by positive energy shift according to the supercell calculations \cite{Song2024AFM}. The AHC ($\sigma_{yz}$) is significantly enhanced by excess Ga and exhibits a maximum at about 0.1 eV, where multiple Weyl points have been identified in Fig.~\ref{fig5}e.

Figure~\ref{fig6}d compares the calculated maximum values of the AHC with the measured ones at 300 K. Due to numerical difficulties, the calculations should only be trusted on a qualitative level. Nevertheless, we find a good agreement in the overall tendency of their magnitudes \cite{Kuroda2017NM,Chen2021NC,Song2024AFM}. The maximum AHC is much higher in Mn$_3$Ga, which, compared to Mn$_3$Ge and Mn$_3$Sn, may be ascribed to the multiple Weyl points above the Fermi energy, suggesting that valence tuning is a potential way to maximize the AHC. For Mn$_3$Ga, this predicts that even higher AHC may be achieved upon further electron doping.

To summarize, we have performed systematic electronic band structure calculations of the kagome altermagnets Mn$_3X$ ($X=$ Sn, Ge, Ga) using DFT+DMFT and found very different results compared to previous DFT calculations. Our work provides a unified framework for understanding the magnetic, electronic, and topological properties across this material family, and captures key experimental observations such as the ARPES spectra and anomalous Hall conductivity without ad hoc adjustments. We also identify the valence environment as a critical factor governing the AHC magnitude. These establish a coherent and material-specific understanding of Mn$_3$X and highlight the importance of correct treatment of electronic correlations beyond DFT for magnetic topological systems.\\

\vspace{20pt}
\noindent
\textbf{Methods}

{\color{black}\noindent \textbf{DFT calculations}}\\
\noindent The first-principles DFT calculations were performed by using the full-potential linearized augmented plane-wave method implemented in the WIEN2k package\cite{2014WIEN2k}. The Muffin-tin radii ($R_{\text{MT}}$) were set to 2.5 a.u. for Mn and Sn, and the cutoff parameter of the basis was chosen such that $R_{\text{MT}}K_{\text{max}}= 8.5$. We took 10000 ${\bf k}$-points in the whole Brillouin zone and used the generalized-gradient approximation with the Perdew-Burke-Ernzerhof (GGA-PBE) exchange-correlation potential \cite{Perdew1996PRL}.\\

{\color{black}\noindent \textbf{DFT+DMFT calculations}}\\
Non-spin-polarized DFT calculations with SOC were first performed using the full potential code WIEN2k \cite{2014WIEN2k}. Cluster DMFT calculations were then carried out for the noncollinear AFM order of Mn$_3$Sn, Mn$_3$Ge, and Mn$_3$Ga, but only the onsite self-energies were considered to avoid the negative sign problem as performed in Ref. \cite{Xu2022nQMa}. The directions of the spin polarization of local orbitals on each Mn-ion were fixed according to experimental observation illustrated in Fig.~\ref{fig1}. The magnitudes of the spin moments were determined self-consistently and found to differ slightly on inequivalent Mn-ions. Only their averages are shown in the figures for simplicity. Both the Hubbard $U$ and the Hund's rule coupling $J$ were added to all five Mn-3$d$ orbitals. The hybridization expansion continuous-time quantum Monte Carlo method (CTQMC) was chosen as the DMFT impurity solver \cite{Haule2007PRB}, and the spectral functions were obtained using the maximum entropy method for analytic continuation of the self-energies \cite{Jarrell1996PR}. The AHC is calculated using the Kubo formula without vertex correction \cite{Khurana1990PRL}
    \begin{equation}
    \begin{split}
    {\rm Re}\sigma_{\alpha \beta}
    =&
    \frac{i\hbar e^2}{V}
    \sum_{\mathbf{k}} \int_{-\infty}^{\infty} d \omega_{1} \int_{-\infty}^{\infty} d \omega_2  
    \left[ f(\omega_1) - f(\omega_2) \right]\\
    &\times {\rm Tr} \left[ V_{\mathbf{k}}^{\alpha} A_{\mathbf{k}}(\omega_1) V_{\mathbf{k}}^{\beta} A_{\mathbf{k}}(\omega_2) \right] 
    \frac{1}{(\omega_2-\omega_1+i\eta)^2},
    \end{split}
\end{equation}
where $\alpha$, $\beta$ are direction indices, $V$ is the volume, $A_{\mathbf{k}}(\omega)$ is the spectral function:
\begin{equation}
    A_{\mathbf{k}}(\omega) = \frac{G_{\mathbf{k}}^{\dagger}(\omega) - G_{\mathbf{k}}(\omega)} {2 \pi i},
\end{equation}
$V_{\mathbf{k}}^{\alpha}$ is the $\alpha$-component of the velocity vector with element in Kohn-Sham states:
\begin{equation}
    V_{\mathbf{k},n_1 n_2}^{\alpha} 
    =
    -\frac{i\hbar}{m} \langle \psi_{\mathbf{k},n_1} | \nabla_\alpha | \psi_{\mathbf{k},n_2} \rangle,
\end{equation} 
$f(\omega) = 1/(e^{\beta \omega} +1)$ is the Fermi distribution function, and the trace is over all band indices. $\eta$ is a small positive infinitesimal. The integration is done over the energy interval $[-3,\,3]$ eV with a fine grid spacing of 1 meV on a $14 \times 14 \times 15$ $\mathbf{k}$-mesh in the Brillouin zone. \\

\vspace{20pt}
\noindent        
\textbf{Acknowledgements}
\noindent We thank Satoru Nakatsuji, Kenta Kuroda, and Takeshi Kondo for generously sharing their ARPES data. This work was supported by the National Key Research and Development Program of China (Grant No. 2022YFA1402203), the National Natural Science Foundation of China (NSFC Grant No. 11974397), and the Strategic Priority Research Program of the Chinese Academy of Sciences (Grant No. XDB33010100).

\vspace{20pt}
\noindent  
\textbf{Code availability}
\noindent We have used the WIEN2k \cite{2014WIEN2k} and DFT+eDMFT codes \cite{Haule2010PRB} to generate the data. WIEN2k is a commercially available software package and DFT+eDMFT can be obtained from http://hauleweb.rutgers.edu.

\end{document}


\title{Correlated topological band structures of the kagome altermagnets Mn$_3X$ ($X=$ Sn, Ge, Ga)}

\author{Yingying Cao}
\affiliation{Beijing National Laboratory for Condensed Matter Physics and
Institute of Physics, Chinese Academy of Sciences, Beijing 100190, China}
\affiliation{School of Physical Sciences, University of Chinese Academy of Sciences, Beijing 100049, China}
\affiliation{Institute of Quantum Materials and Physics, Henan Academy of Sciences, Zhengzhou 450046, China}
\author{Yuanji Xu}
\affiliation{Institute for Applied Physics, University of Science and Technology Beijing, Beijing 100083, China}
\author{Yi-feng Yang}
\email[]{yifeng@iphy.ac.cn}
\affiliation{Beijing National Laboratory for Condensed Matter Physics and
Institute of Physics, Chinese Academy of Sciences, Beijing 100190, China}
\affiliation{School of Physical Sciences, University of Chinese Academy of Sciences, Beijing 100049, China}
\affiliation{Songshan Lake Materials Laboratory, Dongguan, Guangdong 523808, China}

\maketitle
Here we present more details on DFT and DFT+DMFT calculations and their comparisons with ARPES.

\subsection{Comparison of DFT and ARPES in Mn$_3$Sn}
\begin{figure}[b]
        \includegraphics[width=0.85\textwidth]{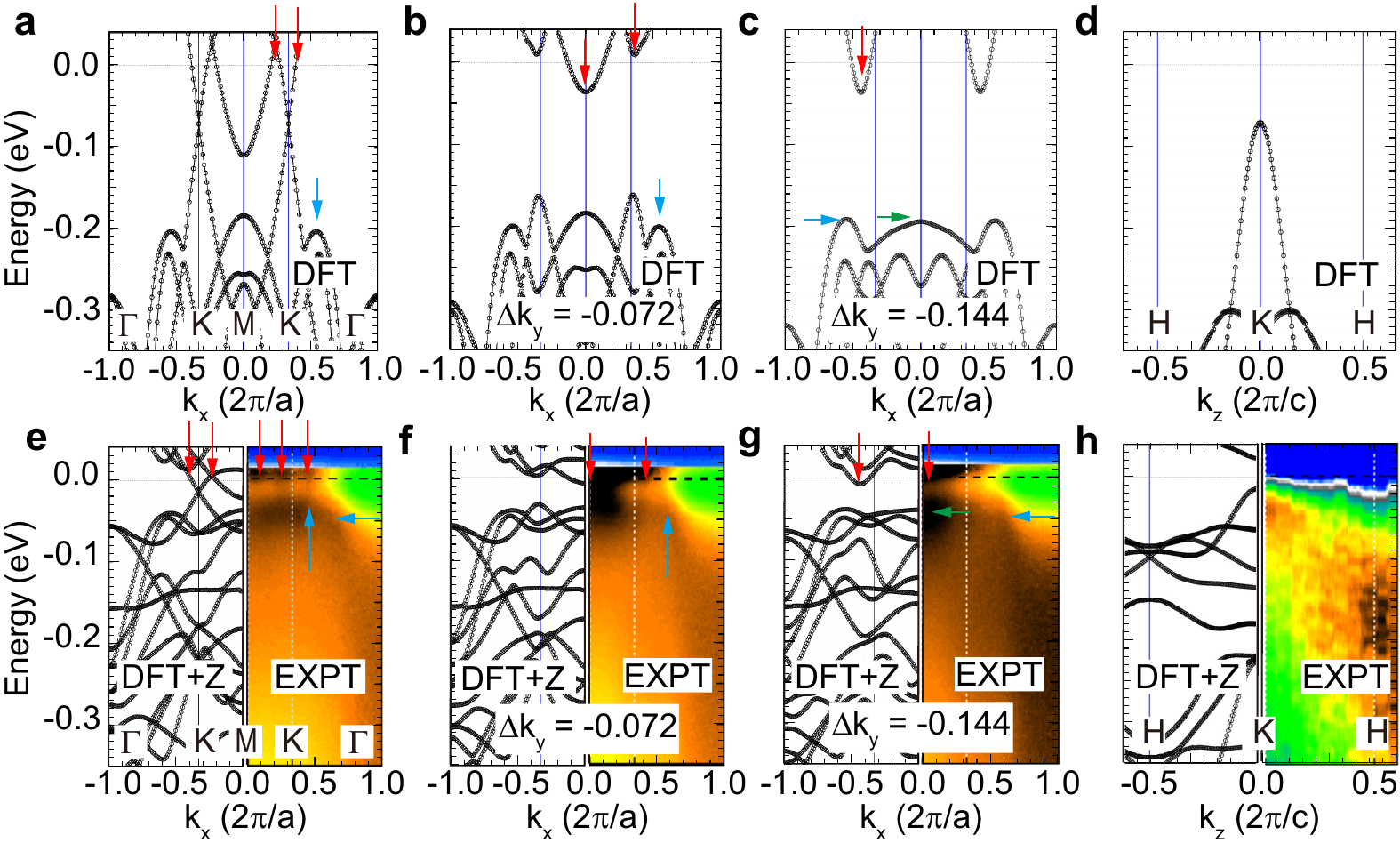}
        \caption{\textbf{Comparison of DFT and ARPES (EXPT) band structures along several paths in the Brillouin zone.} \textbf{a, e} The high-symmetry path $\Gamma$-K-M-K-$\Gamma$; \textbf{b, f} The $k_x$ direction away from the high-symmetry path by $\Delta k_{y} = -0.072$; \textbf{c, g} The $k_x$ direction away from the high-symmetry path by $\Delta k_{y} = -0.144$; \textbf{d, h} The high-symmetry path H-K-H. \textbf{a-d} show the DFT band structures without renormalization, and \textbf{e-h} compare the renormalized DFT band structures (left) and the ARPES data (right) \cite{Kuroda2017NM}.}\label{figS1}
\end{figure}

Figure \ref{figS1} compares the ARPES data \cite{Kuroda2017NM} and the DFT band structures without (a-d) and with (e-h left) artificial renormalization for the AFM1 configuration of Mn$_3$Sn calculated using the noncollinear magnetic version of the WIEN2k package \cite{Laskowski2004PRB}. As shown in Fig.~\ref{figS1}a and the right part of Fig.~\ref{figS1}e, above the Fermi energy the DFT bands without renormalization seem at first glance to match well with experiment along the high-symmetry path $\Gamma$-K-M , but actually the band top around K is located at different energies, about -0.2 eV in DFT and -0.04 eV in ARPES. The right parts of Figs.~\ref{figS1}f and \ref{figS1}g show the ARPES bands along the $k_x$ direction away from the high-symmetry path by a small shift $\Delta k_y$. The electron pockets around K vanish while that around M remains till $\Delta k_y =$ -0.144 (in units of 2$\pi/a$), which is very different in DFT (Figs.~\ref{figS1}b and \ref{figS1}c) where those around K first vanish at $\Delta k_y =$ -0.072 but reappear at $\Delta k_y =$ -0.144 and the one around M disappears at $\Delta k_y =$ -0.144. In addition, there exist obvious spectral weights around -0.2 eV near H in experiment (right part of Fig.~\ref{figS1}h) but nothing is seen in DFT (Fig.~\ref{figS1}d).

Figures \ref{figS1}e-\ref{figS1}h compare the ARPES data (right) and the DFT band structures renormalized by a factor of 5 (left). Now the bands near the Fermi energy along the K-M path and H seem to agree better, but there are still obvious inconsistencies. First, the renormalization introduces a lot of spectral weights along the K-M cut between -0.3 and -0.1 eV, which are not present in experiment (Fig.~\ref{figS1}e). Second, the position of the electron pocket still differs at $\Delta k_y =$ -0.144 away from the high-symmetry path $\Gamma$-K-M as shown in Fig.~\ref{figS1}g.

\subsection{DFT+DMFT band structures with varying $U$ and $J$ in Mn$_3$Sn}

\begin{figure}[b]
        \includegraphics[width=0.92\textwidth]{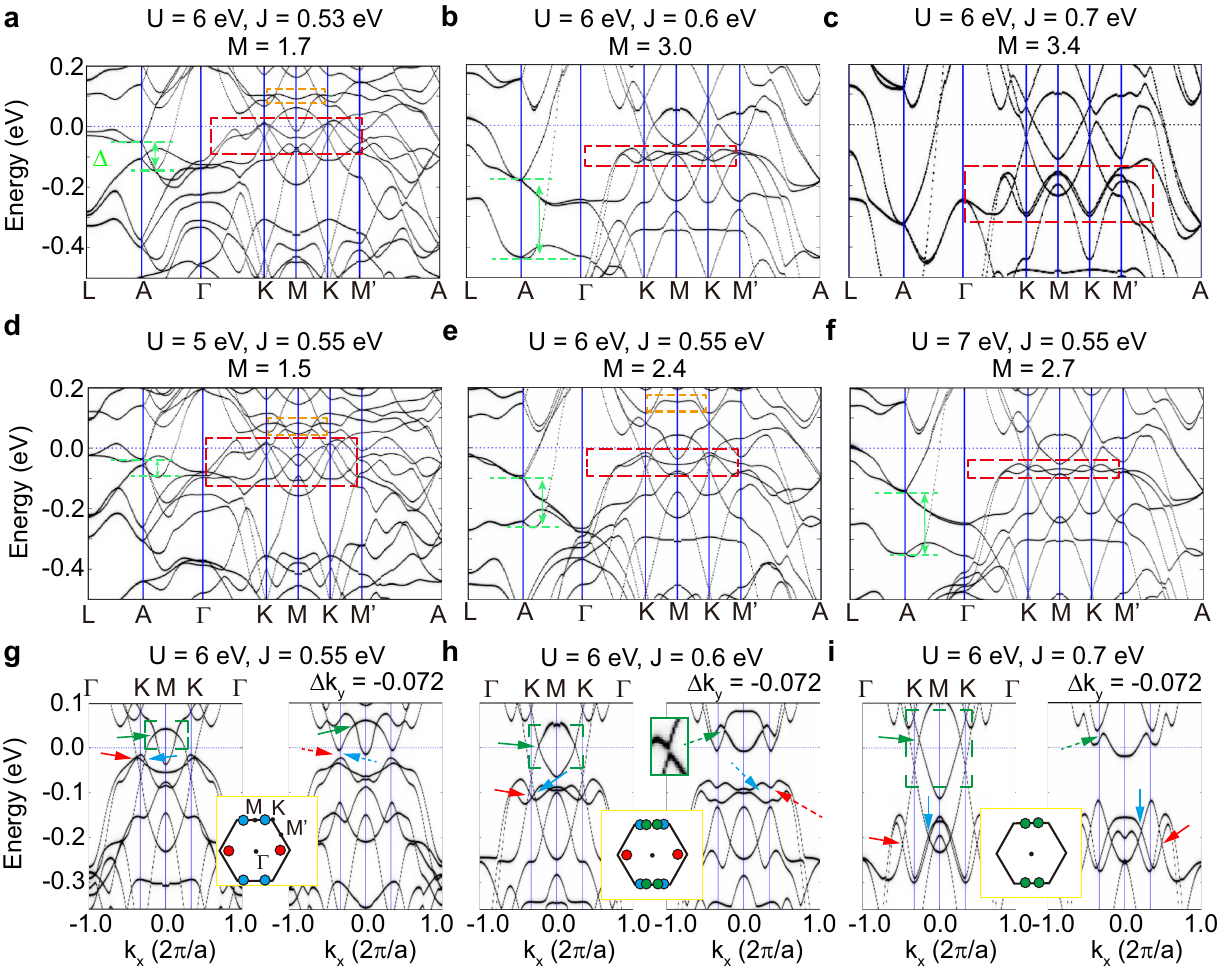}
        \caption{\textbf{Evolution of the DFT+DMFT spectral functions with varying $J$ and $U$ at 300 K.} For clarity, the imaginary part of the self-energy is set to zero to reduce the band broadening. The calculated Mn spin moment is also given for comparison. \textbf{a-c} show the band evolution with varying $J$ for a fixed $U=6\,$eV and \textbf{d-f} show the evolution with $U$ for a fixed $J=0.55\,$eV. The green lines and arrows define the relative shift $\Delta$ between two featured energy bands. The red boxes show the band crossings below the Fermi energy. The orange boxes highlight the M-shaped band above the Fermi energy. \textbf{g-i} show the evolution of the Weyl points in the DFT+DMFT band structures with varying $J$ at $U=6\,$eV. The green boxes show the band crossings above the Fermi energy. The insets show the positions of the Weyl points in the Brillouin zone, as indicated by the arrows of the same color in the spectral function plots. The solid arrows indicate the band crossings and the dashed ones indicate the gaps.}\label{figS2}
\end{figure}

Figure \ref{figS2} shows the DFT+DMFT spectral functions along a high-symmetry path at 300 K for different values of $U$ and $J$. As $J$ increases from 0.53 (a), 0.55 (e), 0.6 (b) to 0.7 eV (c) for a fixed $U=6\,$eV, the bands enclosed in the red box move downwards and the M-shaped band enclosed in the orange box moves up. Such a relative shift of different bands is also seen along A-$\Gamma$ as marked by the green arrows. Similar changes are present in Figs.~\ref{figS2}d-\ref{figS2}f with increasing $U$ for a fixed $J$. A straightforward comparison indicates that the relative band shift $\Delta$ is tuned by the size of the spin moment $M$. In the meanwhile, the bands in the red box get narrowed close to the critical Hund's rule coupling $J_c$, indicating a stronger renormalization near $J_c$ as discussed in Fig.~2d and causing the flat bands below the Fermi energy along the K-M path observed in ARPES \cite{Kuroda2017NM}. 

Thus, the sensitive dependence of the DFT+DMFT band structures on the Hund's rule coupling arises from the rapid variation of the spin moment $M$ above the critical $J_c$. We have shown in the main text that the results at $J=0.55\,$eV for $U=6\,$eV agree well with the ARPES. A slightly larger $J$ can already yield some inconsistency. As shown in Fig.~\ref{figS2}b for $J=0.6\,$eV, the flat bands enclosed in the red box along K-M-K is located at about -0.1 eV rather than -0.04 eV reported in ARPES \cite{Kuroda2017NM}.

More analyses are given in Figs.~\ref{figS2}g-\ref{figS2}i with varying $J$ at $U=6\,$eV and $T=300\,$K. The green boxes mark the band crossings formed by one hole band and one electron band. As $J$ increases, the hole band moves upwards and the electron band moves downwards, but the energy of the crossing points remains almost unchanged. They correspond to the Weyl points in DFT. However, away from the high-symmetry path M-K by a small shift $\Delta k_{y} = -0.072$, the crossings remain for $J=0.55\,$eV but open a gap for $J=0.6$ and 0.7 eV (green arrows) in our DFT+DMFT calculations, indicating that they are not Weyl points at $J=0.55\,$eV. On the other hand, the band crossings inside the red boxes move downwards away from the Fermi energy with increasing $J$. At $J=0.55$ and 0.6 eV, they open a clear gap for $\Delta k_{y} = -0.072$ and form the Weyl points at around -30 meV around K (red and blue arrows) as discussed in the main text. But for $J=0.7\,$eV, they are no longer Weyl points. We thus conclude that the DFT Weyl points only exist above the Fermi energy for large $J$ and evolves into line nodes below $J=0.6\,$eV, while the DFT+DMFT Weyl points appear below the Fermi energy below around $J=0.6\,$eV and move towards the Fermi energy with decreasing $J$.

\subsection{Effect of a small net magnetization in Mn$_3$Sn}
\begin{figure}[h]
    \begin{center}
        \includegraphics[width=0.7\textwidth]{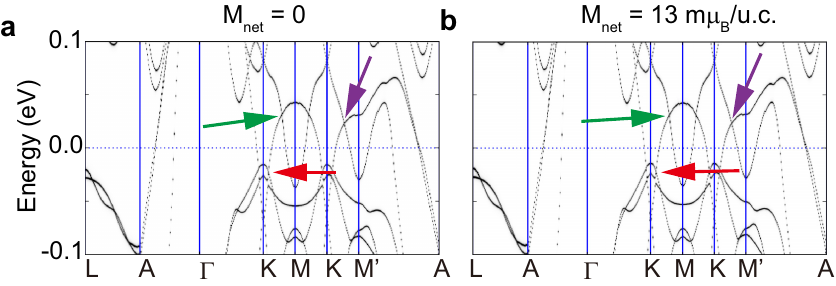}
        \caption{\textbf{Comparison of the DFT+DMFT spectral functions for AFM1 configuration without and with a small net magnetization.} The imaginary part of the self-energy is set to zero for a better view of the band structures. The parameters are $U=6\,$eV, $J=0.55\,$eV, and $T=300\,$K. The net moment is set to 13 m$\mu_{\text B}$/u.c. with the tilted moment from the AFM1 order in \textbf{b} as observed in experiment \cite{Kuroda2017NM}.}\label{figS3}
    \end{center}
\end{figure}

Figure \ref{figS3} compares the DFT+DMFT band structures of AFM1 configuration in Mn$_3$Sn and those with a small net magnetization of 13 m$\mu_{\text B}$/u.c. obtained by rotating all Mn moments towards $a$ axis according to the magnetization experiment reported in Ref. \onlinecite{Kuroda2017NM}. We find that neither the line nodes above the Fermi energy nor the Weyl points below the Fermi energy are affected. Such a small tilting of the spin moments is a negligible perturbation to the electronic topology and thus ignored for simplicity. 

\subsection{More details on the DFT+DMFT band structures in Mn$_3$Ge}
\begin{figure}[h]
    \begin{center}
        \includegraphics[width=0.92\textwidth]{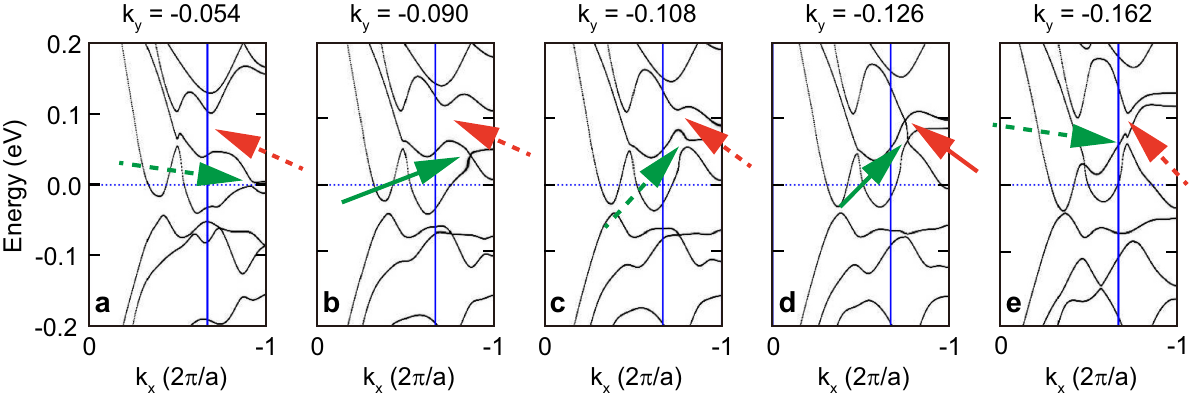}
        \caption{\textbf{DFT+DMFT spectral functions of Mn$_3$Ge along several \textbf{k}-paths parallel to the $x$-axis.} For clarity, the imaginary part of the self-energy is set to zero. The parameters are $U=6\,$eV, $J=0.66\,$eV, and $T=300\,$K. From left to right, the path is shifted by $k_y=-0.054$, -0.09, -0.108, -0.126 and -0.162 away from the high-symmetry $\Gamma$-K-M path.}\label{figS4}
    \end{center}
\end{figure}

Figure \ref{figS4} presents the DFT+DMFT band structures along multiple \textbf{k}-paths parallel to the $x$-axis with varying $k_y$ values in the AFM2 configuration of Mn$_3$Ge. At $k_y$ = -0.126 (Fig. \ref{figS4}d), two band crossings can be identified at 0.06 and 0.09 eV marked by solid green and red arrows. Moving $k_y$ to -0.090 (Fig. \ref{figS4}b) leaves only a single crossing at 0.04 eV (solid green arrow), with the other one getting gapped (red dashed arrow). Further away along the $y$-axis opens the gap for both, as is clearly seen in Figs. \ref{figS4}a, \ref{figS4}c, and \ref{figS4}e. Such systematic gap closure and opening support the presence of Weyl points along the investigated paths.

